\documentclass[aps,preprint]{revtex4}%
\usepackage{amsfonts}
\usepackage{amsmath}
\usepackage{amssymb}
\usepackage{graphicx}%
\begin{document}
\preprint{ }
\title{Werner-like States and Strategic Form of Quantum Games}
\author{Ahmad Nawaz\thanks{email: ahmad@ele.qau.edu.pk}}
\affiliation{National Centre for Physics, QAU Campus Islamabad, Pakistan}
\affiliation{}

\begin{abstract}
We quantize prisoners dilemma, chicken game and battle of sexes to explore the
effect of quantization on their strategic form.The games start with
Werner-like state as an initial state. We show that for the measurement in
entangled basis the strategic forms of these games remain unaffected by
quantization. On the other hand when measurement is performed in product basis
then these games could not retain their strategic forms.

\end{abstract}
\date{\today}
\maketitle

In game theoretic situations two or more rational players compete to maximize
their payoffs by suitable choice of available strategies
\textrm{\cite{dixit,neumann}}. The set of strategies from which unilateral
deviation of any player reduces his/ her payoff is called the Nash Equilibrium
(NE) of the game \cite{nash}.\ In its normal form a game is represented by a
payoff matrix. In order to obtain the strategic form of a game certain
constraints are imposed on the elements of its payoff matrix. Prisoner dilemma
game (PD), for example, is the story of interrogation of two arrested
suspects, Alice and Bob, who have allegedly committed a crime together. Each
of the\ prisoners have to decide whether to confess the crime (to defect $D$)
or to deny the crime (to cooperate $C$) without any communication between
them. According to payoff matrix (\ref{pd-general}),\ if both players receive
$R$ and $U$ for mutual cooperation and mutual defection respectively; and a
cooperator and defector engaged in a contest against each other receive $S$
and $T$ respectively; then the strategic form of PD demands that $T>R>U>S$
\cite{bengt,szabo}. Due these constraints PD takes a form where the rational
reasoning forces each player to defect. As a result $DD$ appears as a NE of
the game with a small payoff $U$ for each player. This NE is not
\textit{Pareto optimal} because the players could have obtained better payoff
$R$\ by playing $C.$ This is referred to as the dilemma of this game.

Chicken game (CG) is another interesting example in this regard. It depicts a
situation in which two players drive their cars straight towards each other.
The first to swerve to avoid the collision $\left(  \text{to cooperate
}C\right)  $ is the loser (chicken) and the one who keeps on driving straight
$\left(  \text{to defect }D\right)  $ is the winner. By assigning $R$ and $U$
to mutual cooperation and defection respectively; $S$ and $T$ to a cooperator
and a defector against each other then the strategic form of CG requires the
constraints on the elements of payoff matrix as $T>R>S>U,$ see payoff matrix
(\ref{pd-general}). Certainly if both players cooperate they can avoid a crash
and none of them will be winner. If one of them steers away $\left(
\text{defects }D\right)  $ he will be loser but will survive but the opponent
will receive the entire honor. If they crash then the cost of both of them
will be higher than the cost of being chicken and the payoff will be lower
\cite{bengt}. There is no dominant strategy and $CD$, $DC$ are two NE in this
game. The dilemma of this game is that $CC$ which is \textit{Pareto optimal}
is not a NE.

The payoff matrix for the Battle of Sexes (BoS) game is of the form
(\ref{bos matrix}). In the usual exposition of this game the players Alice and
Bob are trying to decide a place to spend Saturday evening. Alice wants to
attend Opera while Bob is interested in watching TV at home and both would
prefer to spend the evening together. If $O$ and $T$ represent Opera and TV
respectively and both players receive $\alpha$ and $\beta$\ for playing $O$
and $T$ respectively. They obtain $\sigma$ for strategy pairs $\left(
O,T\right)  $ and $\left(  T,O\right)  .$\ The constraint imposed on the
element of this game is$\ \alpha>\beta>\sigma.$ There exist two NE $(O,O)$ and
$(T,T)$ in the classical form of the game. In absence of any communication
between Alice and Bob, there exists a dilemma as NE $\left(  O,O\right)  $
suits Alice whereas Bob prefers $(T,T).$ As a result both players could end up
with worst payoff $\sigma$\ in case they play mismatched strategies.

The analysis of games in quantum domain helped in resolving such dilemmas. One
of the elegant and foremost step in this direction was by Eisert \textit{el al
}\cite{eisert} to remove dilemma in PD. In this quantization scheme the
strategy space of the players is a two parameter set of $2\times2$ unitary
operators. Starting with maximally entangled initial quantum state the authors
showed that for a suitable quantum strategy the dilemma disappears from the
game. The quantum strategy pair $Q\otimes Q$ appears as a NE with payoffs $R$
for both players and is \textit{Pareto optimal}. They also pointed out that
the quantum strategy $Q$ always wins over all classical strategies. Eisert
\textit{et al} \cite{eisert1} also showed that $Q\otimes Q$ is a unique NE in
CG and is \textit{Pareto optimal}. An experimental demonstration of this
quantization scheme for PD has been achieved on a two qubit nuclear magnetic
resonance (NMR) computer with full range of entanglement parameter $\gamma$
ranging from $0$ to $\frac{\pi}{2} $ \cite{du-2}. It is interesting to note
that these results are in good agreement with theory. Such a type of
demonstration has also been proposed on the optical computer \cite{zhou}. Some
other interesting issues that have been analyzed using this quantization
scheme are, the proof of quantum Nash equilibrium theorem \cite{lee},
evolutionarily stable strategies (ESS) \cite{azhar}, quantum verses classical
player \cite{poit,flitney-1,cheon}, the difference between classical and
quantum correlations \cite{ozdemir,ozdemir1,shimamura} and the\ model of
decoherence in the quantum games \cite{chen,flitney-2}. Eisert \textit{et al}
scheme can easily be implemented to all kinds of $2\times2$ games. A possible
classification of $2\times2$ games has also been given by Huertas-Rosero
\cite{rosero}. Later on, Marinatto and Weber \cite{marinatto} introduced
another interesting and simple scheme for the quantization of non-zero sum
games. They gave Hilbert structure to the strategic spaces of the players.
They also used the maximally entangled initial state and allowed the players
to play their tactics by applying probabilistic choices of unitary operators.
Applying their scheme to Battle of Sexes game they found the strategy for
which both the players have equal payoffs. Marinatto and Weber quantization
scheme gave very interesting results while investigating evolutionarily stable
strategies (ESS) \cite{azhar,azhar1,ahmad02} and in the analysis of repeated
games \cite{azhar2} etc.

In our earlier work we quantized PD and CG to explore the role of quantum
discord in quantum games \cite{nawazdiscord}. To establish this connection we
use Werner-like state as an initial state of the game. We showed that the
dilemma in both PD and CG can be resolved by separable states with non-zero
quantum discord.\textrm{\ }Recently we find that the strategic form of
quantized PD depends upon entanglement of initial quantum state as well as on
the type of measurement basis (entangled or product) \cite{nawazstrategic}.
For both type of measurements there exist respective cutoff values of
entanglement of initial quantum state up to which strategic form of game
remains intact. Beyond these cutoffs the quantized PD behaves like chicken
game up to another cutoff value. Here we show if a quantum game starts with
Werner-like state as an initial quantum state then the strategic form of
quantized game remains intact. Quantizing PD, CG and BoS we show that when
measurement is performed in entangled basis the strategic form of quantized
games is unaffected$.$ However when the measurement is performed in product
basis then PD, CG and BoS lose their strategic form after quantization. This
result high lights the fact that despite being nonlocal, for certain range of
parameter $p,$ when shared between two parties then Werner states behave as a
powerful\ resource in comparison to classical randomness
\textrm{\cite{preskil}}

$%
\begin{array}
[c]{c}
\end{array}
$%
\begin{gather}%
\begin{array}
[c]{ccc}%
\text{ \ \ \ \ \ \ \ \ \ \ } & \text{Bob} &
\end{array}
\nonumber\\%
\begin{array}
[c]{cccc}%
\text{ \ \ \ \ \ \ \ \ \ \ } & \text{ \ \ \ }C & \text{ \ \ \ }D & \text{\ }%
\end{array}
\nonumber\\%
\begin{array}
[c]{c}%
\text{Alice}%
\end{array}%
\begin{array}
[c]{c}%
C\\
D
\end{array}
\left[
\begin{array}
[c]{cc}%
\left(  R,R\right)  & \left(  S,T\right) \\
\left(  T,S\right)  & \left(  U,U\right)
\end{array}
\right] \label{pd-general}\\
\nonumber\\
\text{Matrix 1: The constraints for PD are }T>R>U>S\nonumber\\
\text{and for }T>R>S>U\text{ for CG.}\nonumber
\end{gather}

\begin{align}
&
\begin{array}
[c]{c}%
\text{{\large Alice}}%
\end{array}
\overset{%
\begin{array}
[c]{c}
\end{array}
}{%
\begin{array}
[c]{c}%
O\\
T
\end{array}
}\overset{%
\begin{array}
[c]{c}%
\text{{\large Bob}}\\%
\begin{array}
[c]{cc}%
\text{ }O\text{ \ \ \ \ \ \ } & T
\end{array}
\end{array}
}{\left[
\begin{array}
[c]{cc}%
(\alpha,\beta) & (\sigma,\sigma)\\
(\sigma,\sigma) & (\beta,\alpha)
\end{array}
\right]  }\label{bos matrix}\\
&
\begin{array}
[c]{c}
\end{array}
\nonumber\\
\text{Matrix 2}  & \text{: For BoS it is required that }\alpha>\beta
>\sigma.\nonumber
\end{align}
Before investigating the role of Werner like states in quantum games we
present a brief introduction to these states following Refs.
\cite{werner,munro,ghosh,wei}. Werner states are linear combination of a
maximally entangled and a maximally mixed state. Their entanglement and
nonlocality depends upon a parameter $p$ with values lying in the range $0$
$\leq p\leq1$. For $0<p\leq\frac{1}{3}$ they are separable,\ for $\frac{1}{3}$
$<p\leq\frac{1}{\sqrt{2}}$ entangled but not nonlocal and for the range
$\frac{1}{\sqrt{2}}<p<1$ they become inseparable and nonlocal \cite{werner}.
This behavior is in contrast with their well known separability at $p\leq
\frac{1}{3}.$A two-qubit Werner like state is of the form
\begin{equation}
\rho_{in}=p\left\vert \phi^{+}\right\rangle \left\langle \phi^{+}\right\vert
+\frac{\left(  1-p\right)  }{4}I\otimes I\label{state in}%
\end{equation}
where $\left\vert \phi^{+}\right\rangle $ $=\frac{\left\vert 00\right\rangle
+i\left\vert 11\right\rangle }{\sqrt{2}}$ is standard Bell state.

Next we quantize a\ game with a general payoff matrix given by%

\begin{equation}
\text{{\large Alice }}%
\begin{array}
[c]{c}%
A_{1}\\
A_{2}%
\end{array}
\overset{}{\overset{%
\begin{array}
[c]{c}%
\text{{\large Bob}}\\%
\begin{array}
[c]{cc}%
B_{1}\text{\ \ \ \ } & B_{1}%
\end{array}
\end{array}
}{\left[
\begin{array}
[c]{cc}%
\left(  \$_{00}^{A},\$_{00}^{B}\right)  & \left(  \$_{01}^{A},\$_{01}%
^{B}\right) \\
\left(  \$_{10}^{A},\$_{10}^{B}\right)  & \left(  \$_{11}^{A},\$_{11}%
^{B}\right)
\end{array}
\right]  }} \label{payoff-matrix-general}%
\end{equation}
using Werner like state (\ref{state in}) as an initial quantum state. The
strategy of each of the\ players is represented by the unitary operator
$U_{i}$\ given as\emph{\ }
\begin{equation}
U_{i}=\cos\frac{\theta_{i}}{2}R_{i}+\sin\frac{\theta_{i}}{2}C_{i},\text{
\ \ \ } \label{strategy}%
\end{equation}
where $i=1$\ or $2$\ and $R_{i}$, $C_{i}$\emph{\ }are the unitary operators
defined as%
\begin{align}
R_{i}\left\vert 0\right\rangle  &  =e^{i\phi_{i}}\left\vert 0\right\rangle
,\text{ \ \ }R_{i}\left\vert 1\right\rangle =e^{-i\phi_{i}}\left\vert
1\right\rangle ,\nonumber\\
C_{i}\left\vert 0\right\rangle  &  =-\left\vert 1\right\rangle ,\text{
\ \ \ \ \ }C_{i}\left\vert 1\right\rangle =\left\vert 0\right\rangle .
\label{operators}%
\end{align}
Here we restrict our treatment to two parameter set of strategies for
mathematical simplicity in accordance with Ref. \cite{eisert}.\emph{\ }After
the application of the strategies, the initial state given by Eq.
(\ref{state in}) transforms into
\begin{equation}
\rho_{f}=(U_{1}\otimes U_{2})\rho_{in}(U_{1}\otimes U_{2})^{\dagger}.
\label{final}%
\end{equation}
The payoff operators for Alice and Bob are%

\begin{align}
P^{A} &  =\$_{00}^{A}P_{00}+\$_{11}^{A}P_{11}+\$_{01}^{A}P_{01}+\$_{10}%
^{A}P_{10},\nonumber\\
P^{B} &  =\$_{00}^{B}P_{00}+\$_{11}^{B}P_{11}+\$_{01}^{B}P_{01}+\$_{10}%
^{B}P_{10},\label{pay-operator}%
\end{align}
where
\begin{subequations}
\begin{align}
P_{00} &  =\left\vert \psi_{00}\right\rangle \left\langle \psi_{00}\right\vert
\text{, \ }\left\vert \psi_{00}\right\rangle =\cos\frac{\delta}{2}\left\vert
00\right\rangle +i\sin\frac{\delta}{2}\left\vert 11\right\rangle
,\label{oper 1}\\
P_{11} &  =\left\vert \psi_{11}\right\rangle \left\langle \psi_{11}\right\vert
,\text{ \ }\left\vert \psi_{11}\right\rangle =\cos\frac{\delta}{2}\left\vert
11\right\rangle +i\sin\frac{\delta}{2}\left\vert 00\right\rangle
,\label{oper 2}\\
P_{10} &  =\left\vert \psi_{10}\right\rangle \left\langle \psi_{10}\right\vert
\text{, \ }\left\vert \psi_{10}\right\rangle =\cos\frac{\delta}{2}\left\vert
10\right\rangle -i\sin\frac{\delta}{2}\left\vert 01\right\rangle
,\label{oper 3}\\
P_{01} &  =\left\vert \psi_{01}\right\rangle \left\langle \psi_{01}\right\vert
\text{, \ }\left\vert \psi_{01}\right\rangle =\cos\frac{\delta}{2}\left\vert
01\right\rangle -i\sin\frac{\delta}{2}\left\vert 10\right\rangle
,\label{oper 4}%
\end{align}
with\emph{\ }$\delta\in\left[  0,\frac{\pi}{2}\right]  $ being the
entanglement of the measurement basis. Above payoff operators reduce to that
of Eisert's scheme for $\delta$ equal to $\gamma,$ which represents the
entanglement of the initial state \cite{eisert}. For $\delta=0$ above
operators transform into that of Marinatto and Weber's scheme \cite{marinatto}%
. The payoffs for the players are calculated as
\end{subequations}
\begin{align}
\$_{A}(\theta_{1},\phi_{1},\theta_{2},\phi_{2}) &  =\text{Tr}(P^{A}\rho
_{f})\text{,}\nonumber\\
\$_{B}(\theta_{1},\phi_{1},\theta_{2},\phi_{2}) &  =\text{Tr}(P^{B}\rho
_{f}),\label{payoff-generalized}%
\end{align}
where Tr represents the trace of a\emph{\ }matrix. Using Eqs. (\ref{final}),
(\ref{pay-operator}) and (\ref{payoff-generalized}) the payoffs for players
$j=A,B$ are obtained as%
\begin{align}
\$_{j}(\theta_{1},\phi_{1},\theta_{2},\phi_{2}) &  =\$_{00}^{j}\text{Tr}%
(P_{00}\rho_{f})+\$_{01}^{j}\text{Tr}(P_{01}\rho_{f})+\$_{10}^{j}%
\text{Tr}(P_{10}\rho_{f})+\$_{11}^{j}\text{Tr}(P_{11}\rho_{f})\nonumber\\
& \label{payoffs}%
\end{align}
where we have defined
\begin{subequations}
\label{a}%
\begin{align}
\text{Tr}(P_{00}\rho_{f}) &  =p\left[  \left\{  1-\sin^{2}\left(  \phi
_{1}+\phi_{2}\right)  \sin\delta\right\}  \cos^{2}\frac{\theta_{1}}{2}\cos
^{2}\frac{\theta_{2}}{2}+\right.  \nonumber\\
&  \frac{\left(  \sin\delta-1\right)  }{2}\left\{  \cos^{2}\frac{\theta_{1}%
}{2}+\cos^{2}\frac{\theta_{2}}{2}-\frac{1}{2}\sin\theta_{1}\sin\theta_{2}%
\sin\left(  \phi_{1}+\phi_{2}\right)  \right\}  -\nonumber\\
&  \left.  \frac{\sin\delta}{2}\right]  +\frac{1+p}{4}\label{tr}%
\end{align}%
\end{subequations}
\begin{align}
\text{Tr}(P_{01}\rho_{f}) &  =p\left[  \frac{1+\cos2\phi_{1}\sin\delta}{2}%
\cos^{2}\frac{\theta_{1}}{2}\sin^{2}\frac{\theta_{2}}{2}+\frac{1-\cos2\phi
_{2}\sin\delta}{2}\sin^{2}\frac{\theta_{1}}{2}\cos^{2}\frac{\theta_{2}}%
{2}\right.  +\nonumber\\
&  \left.  \frac{\left(  -1+\sin\delta\right)  \sin\phi_{1}\cos\phi
_{2}-\left(  1+\sin\delta\right)  \cos\phi_{1}\sin\phi_{2}}{4}\sin\theta
_{1}\sin\theta_{2}\right]  +\frac{1-p}{4}\nonumber\\
& \label{trb}%
\end{align}%
\begin{align}
\text{Tr}(P_{10}\rho_{f}) &  =p\left[  \frac{1-\cos2\phi_{1}\sin\delta}{2}%
\cos^{2}\frac{\theta_{1}}{2}\sin^{2}\frac{\theta_{2}}{2}+\frac{1+\cos2\phi
_{2}\sin\delta}{2}\sin^{2}\frac{\theta_{1}}{2}\cos^{2}\frac{\theta_{2}}%
{2}\right.  -\nonumber\\
&  \left.  \frac{\left(  1+\sin\delta\right)  \sin\phi_{1}\cos\phi_{2}+\left(
1-\sin\delta\right)  \cos\phi_{1}\sin\phi_{2}}{4}\sin\theta_{1}\sin\theta
_{2}\right]  +\frac{1-p}{4}\nonumber\\
& \label{trc}%
\end{align}%
\begin{align}
\text{Tr}(P_{11}\rho_{f}) &  =p\left[  \left\{  1-\cos^{2}\left(  \phi
_{1}+\phi_{2}\right)  \sin\delta\right\}  \cos^{2}\frac{\theta_{1}}{2}\cos
^{2}\frac{\theta_{2}}{2}+\right.  \nonumber\\
&  \left.  \frac{\left(  \sin\delta+1\right)  }{2}\left\{  \sin^{2}%
\frac{\theta_{1}}{2}\sin^{2}\frac{\theta_{2}}{2}+\frac{1}{2}\sin\theta_{1}%
\sin\theta_{2}\sin\left(  \phi_{1}+\phi_{2}\right)  \right\}  \right]
+\nonumber\\
&  +\frac{1-p}{4}\label{trd}%
\end{align}
In the framework of our generalized quantization scheme \cite{nawaz}
measurement can be performed either using entangled basis $\left(
\delta=\frac{\pi}{2}\right)  $ or product basis $\left(  \delta=0\right)  $.
Next we discuss both these cases one by one.\pagebreak

\textbf{Case 1: Measurement in entangled basis}

For the measurement in entangled basis with the help of Eq. (\ref{payoffs})
the payoffs for players become
\begin{align}
\$_{j}(\theta_{1},\phi_{1},\theta_{2},\phi_{2})  &  =p\left[  \$_{00}%
^{j}\left(  \cos^{2}\left(  \phi_{1}+\phi_{2}\right)  \cos^{2}\frac{\theta
_{1}}{2}\cos^{2}\frac{\theta_{2}}{2}\right)  \right.  +\nonumber\\
&  \$_{01}^{j}\left(  \cos\phi_{1}\cos\frac{\theta_{1}}{2}\sin\frac{\theta
_{2}}{2}-\sin\phi_{2}\sin\frac{\theta_{1}}{2}\cos\frac{\theta_{2}}{2}\right)
^{2}+\nonumber\\
&  \$_{10}^{j}\left(  \sin\phi_{1}\cos\frac{\theta_{1}}{2}\sin\frac{\theta
_{2}}{2}-\cos\phi_{2}\sin\frac{\theta_{1}}{2}\cos\frac{\theta_{2}}{2}\right)
^{2}+\nonumber\\
&  \left.  \$_{11}^{j}\left(  \cos\frac{\theta_{1}}{2}\cos\frac{\theta_{2}}%
{2}\sin\left(  \phi_{1}+\phi_{2}\right)  +\sin\frac{\theta_{1}}{2}\sin
\frac{\theta_{2}}{2}\right)  ^{2}\right] \nonumber\\
&  +\frac{\left(  1-p\right)  }{4}\left(  \$_{00}^{j}+\$_{01}^{j}+\$_{10}%
^{j}+\$_{11}^{j}\right)  \label{payoff-delta-Pi/2}%
\end{align}
For PD with payoff matrix elements $\$_{00}^{A}=\$_{00}^{B}=3,\$_{01}%
^{A}=\$_{10}^{B}=0,\$_{10}^{A}=\$_{01}^{B}=5$ and $\$_{11}^{A}=\$_{11}^{B}=1$
the above equation reduces to
\begin{align}
\$_{A}(\theta_{1},\phi_{1},\theta_{2},\phi_{2})  &  =p\left[  3\left(
\cos^{2}\left(  \phi_{1}+\phi_{2}\right)  \cos^{2}\frac{\theta_{1}}{2}\cos
^{2}\frac{\theta_{2}}{2}\right)  \right.  +\nonumber\\
&  5\left(  \sin\phi_{1}\cos\frac{\theta_{1}}{2}\sin\frac{\theta_{2}}{2}%
-\cos\phi_{2}\sin\frac{\theta_{1}}{2}\cos\frac{\theta_{2}}{2}\right)
^{2}+\nonumber\\
&  \left.  \left(  \cos\frac{\theta_{1}}{2}\cos\frac{\theta_{2}}{2}\sin\left(
\phi_{1}+\phi_{2}\right)  +\sin\frac{\theta_{1}}{2}\sin\frac{\theta_{2}}%
{2}\right)  ^{2}\right]  +\nonumber\\
&  \frac{9}{4}\left(  1-p\right)  \label{payoffs-pd-a}%
\end{align}%
\begin{align}
\$_{B}(\theta_{1},\phi_{1},\theta_{2},\phi_{2})  &  =p\left[  3\left(
\cos^{2}\left(  \phi_{1}+\phi_{2}\right)  \cos^{2}\frac{\theta_{1}}{2}\cos
^{2}\frac{\theta_{2}}{2}\right)  \right.  +\nonumber\\
&  5\left(  \cos\phi_{1}\cos\frac{\theta_{1}}{2}\sin\frac{\theta_{2}}{2}%
-\sin\phi_{2}\sin\frac{\theta_{1}}{2}\cos\frac{\theta_{2}}{2}\right)
^{2}+\nonumber\\
&  \left.  \left(  \cos\frac{\theta_{1}}{2}\cos\frac{\theta_{2}}{2}\sin\left(
\phi_{1}+\phi_{2}\right)  +\sin\frac{\theta_{1}}{2}\sin\frac{\theta_{2}}%
{2}\right)  ^{2}\right]  +\nonumber\\
&  \frac{9}{4}\left(  1-p\right)  \label{payoffs-pd-b}%
\end{align}
For $p=1$ these results reduce to that of Eisert et al. \cite{eisert} and the
dilemma in game is resolved for players strategies $U(\theta_{1},\phi
_{1},\theta_{2},\phi_{2})=U(0,\frac{\pi}{2},0,\frac{\pi}{2})=Q$ with
$\$_{A}(0,\frac{\pi}{2},0,\frac{\pi}{2})=\$_{B}(0,\frac{\pi}{2},0,\frac{\pi
}{2})=\left(  3,3\right)  $. Next we investigate whether the strategy $Q$ is
NE for $p\neq1.$ Then the NE\ conditions
\begin{align}
\$_{A}(0,\frac{\pi}{2},0,\frac{\pi}{2})-\$_{A}(\theta_{1},\phi_{1},0,\frac
{\pi}{2})  &  \geq0\nonumber\\
\$_{B}(0,\frac{\pi}{2},0,\frac{\pi}{2})-\$_{B}(0,\frac{\pi}{2},\theta_{2}%
,\phi_{2})  &  \geq0 \label{NE}%
\end{align}
give%
\begin{equation}
p\left(  3\sin^{2}\frac{\theta_{1}}{2}+2\cos^{2}\frac{\theta_{1}}{2}\cos
^{2}\phi_{1}\right)  \geq0.
\end{equation}
The above inequality is satisfied for all values of $p\geq0$ showing that the
strategy pair $\left(  Q,Q\right)  $ continues to be NE for all values of
$p>0.$ It shows that although state (\ref{state in}) is not entangled for
$p\leq\frac{1}{3}$ yet when shared between two players it is proved to be a
better resource as compared to classical randomness. On the other hand at
$p=0$ when the initial state becomes maximally mixed the payoffs become
$\frac{9}{4}$ irrespective of players strategies.

\begin{figure}
[h]
\begin{center}
\includegraphics[
height=2.2589in,
width=3.3615in
]%
{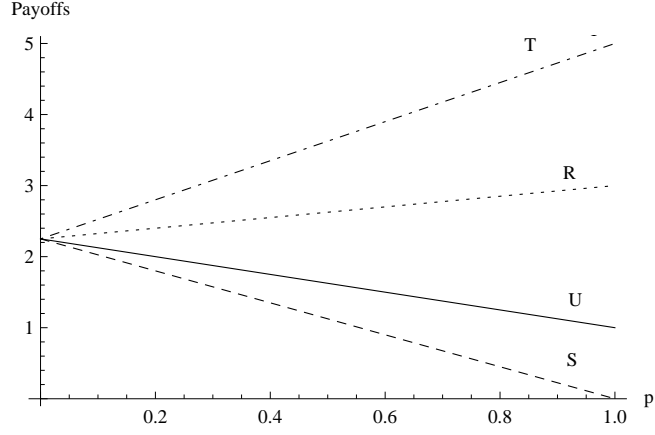}%
\caption{The payoff elements $R,S,$ $T$ and $U$ versus $p.$ It shows that the
constraints required to maintain the strategic form of PD are satisfied for
all values of $p>0.$}%
\label{PD1}%
\end{center}
\end{figure}

Now we investigate whether the quantized PD with $Q\otimes Q$ as NE has the
strategic form like that of PD. Using Eqs. (\ref{payoffs-pd-a},
\ref{payoffs-pd-b}) the elements of payoff matrix of quantized PD are
\begin{equation}
R=\frac{3}{4}p+\frac{9}{4},\text{ }S=\frac{9}{4}-\frac{9}{4}p,\text{ }%
T=\frac{11}{4}p+\frac{9}{4},\text{ }U=\frac{9}{4}-\frac{5}{4}p.\text{
}\label{rstu-entangled}%
\end{equation}
It is easy to see that these payoff elements obey the constraints $T>R>U>S$
for all values of $p>0.$ Therefore we conclude that the strategic form of PD
remains unaffected by quantization if it starts with Werner-like state as an
initial quantum state. The payoff elements (\ref{rstu-entangled}) are shown in
figure (\ref{PD1}) which shows that the payoffs elements obey the constraints
required by PD.%

For CG with payoff matrix elements $\$_{00}^{A}=\$_{00}^{B}=3,\$_{01}%
^{A}=\$_{10}^{B}=1,\$_{10}^{A}=\$_{01}^{B}=4$ and $\$_{11}^{A}=\$_{11}^{B}=0$
the payoffs given in Eq. (\ref{payoff-delta-Pi/2}) become
\begin{align}
\$_{A}(\theta_{1},\phi_{1},\theta_{2},\phi_{2}) &  =p\left[  3\left(  \cos
^{2}\left(  \phi_{1}+\phi_{2}\right)  \cos^{2}\frac{\theta_{1}}{2}\cos
^{2}\frac{\theta_{2}}{2}\right)  \right.  +\nonumber\\
&  \left(  \cos\phi_{1}\cos\frac{\theta_{1}}{2}\sin\frac{\theta_{2}}{2}%
-\sin\phi_{2}\sin\frac{\theta_{1}}{2}\cos\frac{\theta_{2}}{2}\right)
^{2}+\nonumber\\
&  \left.  4\left(  \sin\phi_{1}\cos\frac{\theta_{1}}{2}\sin\frac{\theta_{2}%
}{2}-\cos\phi_{2}\sin\frac{\theta_{1}}{2}\cos\frac{\theta_{2}}{2}\right)
^{2}\right]  +\nonumber\\
&  2\left(  1-p\right)  \label{payoff-chicken-a}%
\end{align}%
\begin{align}
\$_{B}(\theta_{1},\phi_{1},\theta_{2},\phi_{2}) &  =p\left[  3\left(  \cos
^{2}\left(  \phi_{1}+\phi_{2}\right)  \cos^{2}\frac{\theta_{1}}{2}\cos
^{2}\frac{\theta_{2}}{2}\right)  \right.  +\nonumber\\
&  \left(  \sin\phi_{1}\cos\frac{\theta_{1}}{2}\sin\frac{\theta_{2}}{2}%
-\cos\phi_{2}\sin\frac{\theta_{1}}{2}\cos\frac{\theta_{2}}{2}\right)
^{2}+\nonumber\\
&  \left.  4\left(  \cos\phi_{1}\cos\frac{\theta_{1}}{2}\sin\frac{\theta_{2}%
}{2}-\sin\phi_{2}\sin\frac{\theta_{1}}{2}\cos\frac{\theta_{2}}{2}\right)
^{2}\right]  +\nonumber\\
&  2\left(  1-p\right)  \label{payoff-chicken-b}%
\end{align}
With the help of Eqs. (\ref{NE}) the strategy pair $U(\theta_{1},\phi
_{1},\theta_{2},\phi_{2})=U(0,\frac{\pi}{2},0,\frac{\pi}{2})$ will be NE\ of
this game if
\begin{equation}
p\left[  2+\cos^{2}\frac{\theta_{1}}{2}\left(  3\cos^{2}\phi_{1}-2\right)
\right]  \geqslant0.
\end{equation}
The above condition is satisfied for all values of $p\geq0.$ It means that
dilemma can be resolved in CG when the players share the state (\ref{state in}%
) with $p>0$. Furthermore it can be checked by Eqs. (\ref{payoff-chicken-a},
\ref{payoff-chicken-b}) that for $p=0$ the payoffs of the players become $2$,
independent of players decisions.

\begin{figure}
[h]
\begin{center}
\includegraphics[
height=2.2589in,
width=3.3615in
]%
{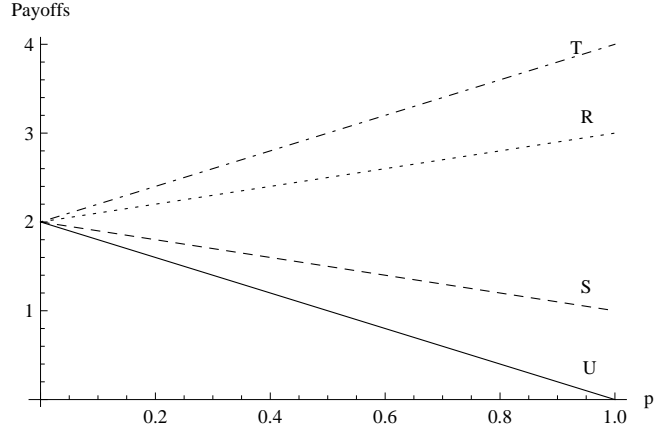}%
\caption{The payoff elements $R,S,$ $T$ and $U$ versus $p.$ It shows that the
constraints required to maintain the strategic form of CG are satisfied for
all values of $p>0.$ }%
\label{CG}%
\end{center}
\end{figure}

Now we investigate the strategic form of quantized CG with $Q\otimes Q$ as a
NE. Using Eqs. (\ref{payoff-chicken-a}, \ref{payoff-chicken-b}) the elements
of payoff matrix of quantized CG become
\begin{equation}
R=p+2,\text{ }S=2-p,\text{ }T=2p+2,\text{ }U=2-2p.\text{ }%
\end{equation}
It is evident that these payoff elements obey the constraints $T>R>S>U$
required for a game to behave like CG for all values of $p>0$. We plot these
payoff elements in figure (\ref{CG}) which shows that the strategic form of CG
is not affected by quantization when it starts with an initial state of the
form of Werner-like states.%

For BoS using payoff matrix elements $\alpha=2,\beta=1$ and $\gamma=0$ the
payoffs given in Eq. (\ref{payoff-delta-Pi/2}) become%
\begin{align}
\$_{A}(\theta_{1},\phi_{1},\theta_{2},\phi_{2})  &  =p\left[  2\left(
\cos^{2}\left(  \phi_{1}+\phi_{2}\right)  \cos^{2}\frac{\theta_{1}}{2}\cos
^{2}\frac{\theta_{2}}{2}\right)  \right.  +\nonumber\\
&  \left.  \left(  \cos\frac{\theta_{1}}{2}\cos\frac{\theta_{2}}{2}\sin\left(
\phi_{1}+\phi_{2}\right)  +\sin\frac{\theta_{1}}{2}\sin\frac{\theta_{2}}%
{2}\right)  ^{2}\right] \nonumber\\
&  +\frac{3\left(  1-p\right)  }{4} \label{payoff a bos}%
\end{align}%
\begin{align}
\$_{B}(\theta_{1},\phi_{1},\theta_{2},\phi_{2})  &  =p\left[  \left(  \cos
^{2}\left(  \phi_{1}+\phi_{2}\right)  \cos^{2}\frac{\theta_{1}}{2}\cos
^{2}\frac{\theta_{2}}{2}\right)  \right.  +\nonumber\\
&  \left.  2\left(  \cos\frac{\theta_{1}}{2}\cos\frac{\theta_{2}}{2}%
\sin\left(  \phi_{1}+\phi_{2}\right)  +\sin\frac{\theta_{1}}{2}\sin
\frac{\theta_{2}}{2}\right)  ^{2}\right] \nonumber\\
&  +\frac{3\left(  1-p\right)  }{4} \label{payoff b bos}%
\end{align}
With the help of the above payoffs we see that
\begin{equation}
\alpha=\frac{5}{4}p+\frac{3}{4},\beta=\frac{1}{4}p+\frac{3}{4},\sigma=\frac
{3}{4}-\frac{3}{4}p \label{element bos}%
\end{equation}
These payoff elements obey the constraints required by a game to behave like
BoS for all values of $p>0$. We plot the payoff elements in figure (\ref{bos})
below. It is clear that after quantization the strategic form of BoS remain
unaffected if it starts with a Werner-like initial quantum state.%

\begin{figure}
[h]
\begin{center}
\includegraphics[
height=2.2425in,
width=3.3615in
]%
{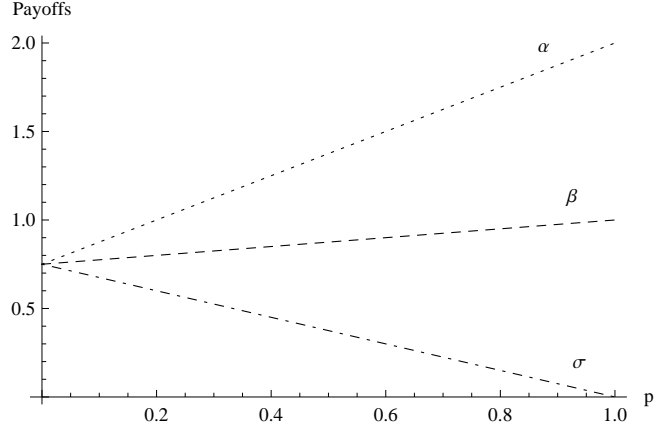}%
\caption{The payoff elements $\alpha,$ $\beta$ and $\gamma$ versus $p.$ It
shows that the constraints required to maintain the strategic form of BoS are
satisfied for all values of $p>0.$}%
\label{BOS}%
\end{center}
\end{figure}

It shows that for quantized versions of PD, CG and BoS the strategic forms the
games remain unaffected if the initial quantum state is Werner-like state.

\textbf{Case 2: Measurement in product basis}

For the measurement performed in product basis (i.e. for $\delta=0$ in Eqs.
(\ref{oper 1} to \ref{oper 4}) ) the Eq. (\ref{payoffs}) reduces to
\begin{align}
\$_{j}(\theta_{1},\phi_{1},\theta_{2},\phi_{2})  &  =\frac{p}{2}\left[
\left(  \$_{00}^{j}+\$_{11}^{j}\right)  \left\{  \cos^{2}\frac{\theta_{1}}%
{2}\cos^{2}\frac{\theta_{2}}{2}+\sin^{2}\frac{\theta_{1}}{2}\sin^{2}%
\frac{\theta_{2}}{2}+\right.  \right. \nonumber\\
&  \left.  \frac{1}{2}\sin\theta_{1}\sin\theta_{2}\sin\left(  \phi_{1}%
+\phi_{2}\right)  \right\}  +\left(  \$_{01}^{j}+\$_{10}^{j}\right)  \left\{
\cos^{2}\frac{\theta_{1}}{2}\sin^{2}\frac{\theta_{2}}{2}\right. \nonumber\\
&  +\left.  \left.  \sin^{2}\frac{\theta_{1}}{2}\cos^{2}\frac{\theta_{2}}%
{2}-\frac{1}{2}\sin\theta_{1}\sin\theta_{2}\sin\left(  \phi_{1}+\phi
_{2}\right)  \right\}  \right]  +\nonumber\\
&  \frac{\left(  1-p\right)  }{4}\left(  \$_{00}^{j}+\$_{01}^{j}+\$_{10}%
^{j}+\$_{11}^{j}\right)  \label{payoffs-delta=0}%
\end{align}
\textrm{For }$p>0$\textrm{\ the above payoffs remain equivalent to the payoffs
obtained by Marinatto and Weber's quantization scheme where the players also
have the option to manipulate the phase }$\phi$\textrm{\ of the given qubit
\cite{marinatto,ma}. However at }$p=0$\textrm{\ when the quantum discord
disappears the payoffs given by Eq. (\ref{payoffs-delta=0}) become average
value of the entries of payoff matrix (\ref{payoff-matrix-general}).}

For PD with payoff matrix elements $\$_{00}^{A}=\$_{00}^{B}=3,\$_{01}%
^{A}=\$_{10}^{B}=0,\$_{10}^{A}=\$_{01}^{B}=5$ and $\$_{11}^{A}=\$_{11}^{B}=1$
the payoff given in Eq. (\ref{payoffs-delta=0}) become
\begin{equation}
R=U=\text{ }\frac{9}{4}-\frac{p}{4},\text{ }S=T=\frac{p}{4}+\frac{9}{4}.\text{
}%
\end{equation}
For CG with payoff matrix elements $\$_{00}^{A}=\$_{00}^{B}=3,\$_{01}%
^{A}=\$_{10}^{B}=1,\$_{10}^{A}=\$_{01}^{B}=4$ and $\$_{11}^{A}=\$_{11}^{B}=0$
using Eq. (\ref{payoffs-delta=0}) we get
\begin{equation}
R=U=\text{ }2-\frac{p}{2},\text{ }S=T=\frac{p}{2}+2.\text{ }%
\end{equation}
Similarly for BoS with payoff matrix elements $\alpha=2,\beta=1$ and
$\gamma=0$ the elements of the quantized payoff matrix become%

\begin{equation}
\alpha=\beta=\frac{3}{4}p+\frac{3}{4},\gamma=\frac{3}{4}-\frac{3}{4}p.
\end{equation}
For all three games it is easy to check that when the measurement is performed
in product basis then the strategic form of the game never remains the same.

In summary we quantized PD, CG and BoS taking Werner-like state as an initial
quantum state to explore the strategic form of their quantized versions. We
performed measurements in entangled and product basis.\emph{\ }For the
measurement in entangled basis we showed that the strategic form of quantized
PD, CG and BoS remains intact for all values of $p>0$.\textrm{\ }This
highlights the fact that despite being nonlocal, for certain range of
parameter $p,$ when shared between two parties these states are a
powerful\ resource in comparison to classical randomness \cite{preskil}. On
the other hand when measurement is performed in product basis then the
strategic form of all the three games does not remain intact.

\end{document}